\documentclass[aps,amsmath,showpacs,amsfonts,10pt]{revtex4}
\usepackage{epsfig,graphicx}
\usepackage[english]{babel}
\usepackage{amsfonts}
\usepackage{amsmath}
\usepackage{latexsym}
\usepackage{graphics,bm}
\usepackage{subfigure}
\usepackage{dcolumn}
\usepackage{bm}
\usepackage{rotating}

\begin{document}

\title{Faraday instability in a two-component Bose Einstein condensate}

\author{Aranya B Bhattacherjee}
\affiliation{Max Planck-Institute f\"ur Physik komplexer Systeme, N\"othnitzer Str.38, 01187 Dresden, Germany }

\begin{abstract}
Motivated by recent experiments on Faraday waves in Bose Einstein condensates (BEC) we investigate the dynamics of two component cigar shaped BEC subject to periodic modulation of the strength of the transverse confinement. It is shown that two coupled Mathieu equations govern the dynamics of the system. We found that the two component BEC in a phase mixed state is relatively more unstable towards pattern formation than the phase segregated state.
\end{abstract}


\maketitle

\section{Introduction}

Faraday waves are generated when the free surface of a fluid layer is subjected to a periodic vertical acceleration \cite{Faraday}. When the acceleration exceeds a threshold value, surface waves appear oscillating at half the forcing frequency. Recently, Faraday waves, which comes under the category of spontaneous pattern formation, a very general phenomenon studied in different fields of nonlinear science,  were observed in a cigar shaped BEC by periodically modulating the radial trap frequency \cite{Engels07}. The radial modulation leads to a periodic modulation of the density of the cloud in time, which in turn leads to a periodic change in the non-linear interactions. This leads to the parametric excitation of longitudinal soundlike waves (Faraday waves) in the direction of weak confinement.
It has been shown theoretically that for a BEC, Faraday instability can be generated either by modulating the scattering length by Feshbach resonance \cite{Stalinus02} or by modulating the trap frequency in the tight confinement direction \cite{Stalinus04}. In both the cases, the dynamics are governed by a Mathieu equation that is typical for parametrically driven systems. Floquet analysis reveals that a series of resonances exists, consisting of a main resonance at half the driving frequency and higher resonance tongues at interger multiples of half the driving frequency . A theoretical analysis of Faraday instability based on a Mathieu-type analysis of the non-polynomial Schr\"odinger equation was given recently \cite{Nicolin07}. From the perspective of phonon number occupation, the Faraday type modulation has been analyzed theoretically \cite{Kagan01}. In general, the nonlinear spatiotemporal dynamics of BECs is attracting increasing interest in recent years, with a major focus on controlling structures like solitons by temporal modulation of the atomic scattering length \cite{Saito}. The evolution of a BEC in a time dependent trap has been addressed earlier by some authors \cite{Juan}.

The aim of the present paper is to analyze Faraday instability in a two-component cigar shaped BEC. The two components could be for example $^{87}Rb$ atoms in two different hyperfine states in different external trapping potentials. A periodic modulation of the radial trap frequencies would give rise to Faraday instability in both the components. The dynamics of this system is now governed by two coupled Mathieu equations. If $\Omega_{1}$ and $\Omega_{2}$ are the natural frequencies of the two components then the parametric resonances will occur not only at $2 \Omega_{1}/n$ and $2 \Omega_{2}/n$ ($n$ is an integer) but also at the combination frequencies ($|\pm \Omega_{1} \pm \Omega_{2}|/n$). The behaviour of this system would then be similar to spontaneous pattern formation in miscible and immiscible binary classical liquids \cite{Huke}. 

An important question is the stability of such Faraday waves in binary mixtures because depending on the atom-atom interaction strengths $g_{ii}$ (intra species interactions) and $g_{ij},i \neq j$ (inter species interaction), binary mixtures can be phase separated ($g_{11}g_{22} < g^{2}_{12}$) or phase mixed ($g_{11}g_{22} > g^{2}_{12}$) \cite{Pethick}. The nonlinear interaction can be conveniently manipulated by Feshbach resonances \cite{Inouye}. The purpose of the present paper is to analyze the influence of the interactions on the stability of the parametric excitations that are generated as a result of periodic modulation of the radial confinements of the two cigar shaped elongated components of the BEC held in a magnetic trap. In contrast to binary liquids, the interactions in the present system are tunable and one can in the same experiment crossover from one state to the other. In particular, we show that a phase mixed state is more unstable towards parametric excitations than a phase separated state.

\begin{figure}[t]
\hspace{-2.0cm}
\includegraphics{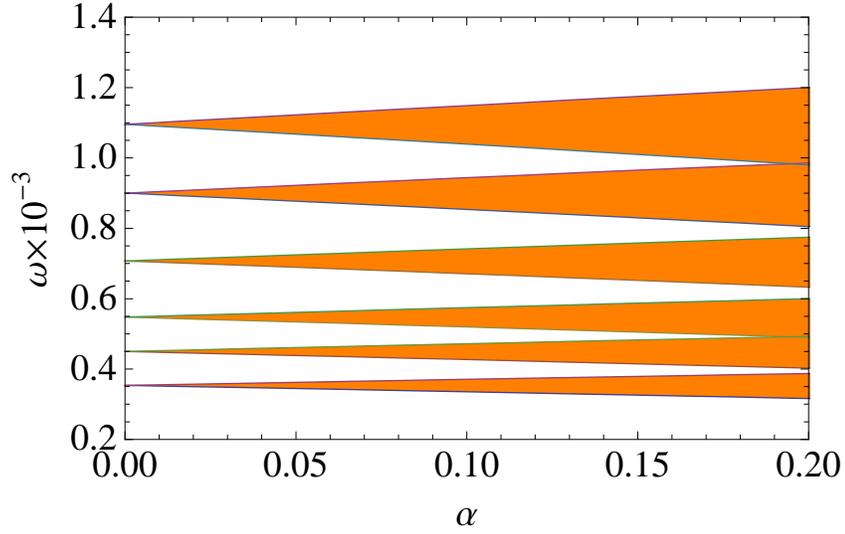} 
\caption{Resonance tongues of the parametric instability for the dissipationless phase separated state ($g_{1}=0.5, g_{2}=0.75$). Shaded domains indicate where the two-component system is unstable. Starting from the lowest tongue, the resonances are ${\Omega_{2},\dfrac{\Omega_{1}+\Omega_{2}}{2}},\Omega_{1}$ and ${2\Omega_{2},\Omega_{1}+\Omega_{2}},2\Omega_{1}$.}
\label{1}
\end{figure}

\begin{figure}[t]
\hspace{-2.0cm}
\includegraphics{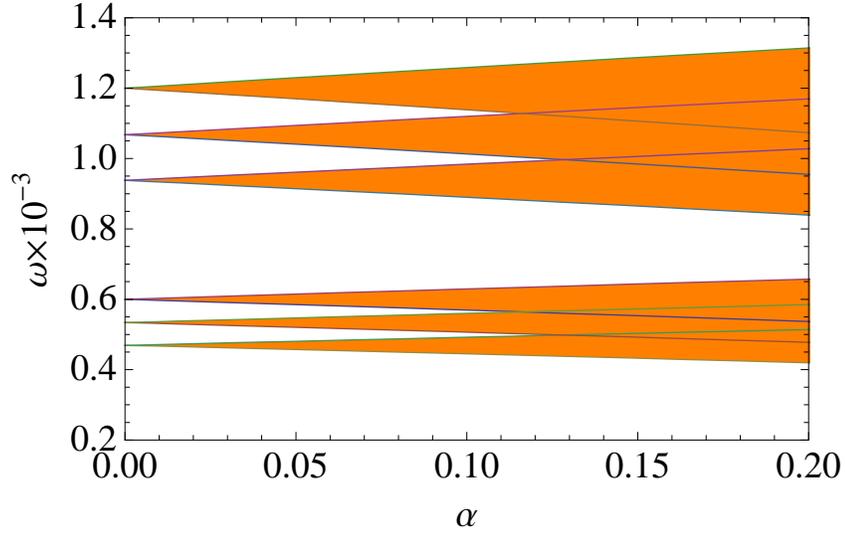} 
\caption{Resonance tongues of the parametric instability for the dissipationless phase mixed state ($g_{1}=1.25, g_{2}=1.5$). Shaded domains indicate where the two-component system is unstable. Starting from the lowest tongue, the resonances are ${\Omega_{2},\dfrac{\Omega_{1}+\Omega_{2}}{2}},\Omega_{1}$ and ${2\Omega_{2},\Omega_{1}+\Omega_{2}},2\Omega_{1}$. Clearly compared to Figure 1, nearby resonance tongues are overlapping and thereby the stable regions around the resonance tongues are reduced indicating that a lower value of the perturbation is sufficient to induce a Faraday instability in the phase mixed state as compared to a phase segregated state.}
\label{2}
\end{figure}

\section{The model}

We consider a situation in which we have a two component BEC (bosonic atoms of the same isotope but having different internal spin states, eg. $^{87}Rb$ atoms in hyperfine states $F=2, m_{F}=2$ and $F=1, m_{F}=-1$). The starting point of our analysis is the coupled Gross Pitaeviskii (GP) equations for a two component confined BEC:

\begin{equation}
\imath \hbar \dfrac{\partial \psi_{1}}{\partial t}=\left\lbrace -\dfrac{\hbar^{2}}{2m} \nabla^{2}+V_{1}(\vec r,t)+g_{11}N_{1}|\psi_{1}|^{2}+g_{12}N_{2}|\psi_{2}|^{2}-\mu_{1}\right\rbrace \psi_{1} 
\end{equation}

\begin{equation}
\imath \hbar \dfrac{\partial \psi_{2}}{\partial t}=\left\lbrace -\dfrac{\hbar^{2}}{2m} \nabla^{2}+V_{2}(\vec r,t)+g_{22}N_{2}|\psi_{2}|^{2}+g_{21}N_{1}|\psi_{1}|^{2}-\mu_{2}\right\rbrace \psi_{2} 
\end{equation}

Here $\psi_{i}$, with $i=1,2$ is the effective wavefunction of the $i^{th}$ condensate, with mass $m$ and chemical potential $\mu_{i}$. The interaction between the $i^{th}$ condensate atoms is specified by $g_{ii}=\dfrac{4 \pi \hbar^{2} a_{ii}}{m}$ and that between $1$ and $2$ by $g_{12}=g_{21}=\dfrac{4 \pi \hbar^{2} a_{12}}{m}$. Here $a_{ii}$ is the intraspecies $s$-wave scattering length and $a_{12}=a_{21}$ is the interspecies scattering length. $N_{i}$ is the number of particles of the $i^{th}$ species. Here $a_{ii}(a_{ij})>0$ for a repulsive BEC, which we consider. $V_{i}(\vec r,t)$ is the trapping potential of the $i^{th}$ component. The wavefunctions are normalized according to $\int |\psi_{i}|^{2} d\vec r=1$ in the absence of damping. Here we are ignoring damping and by doing so, we do not loose the essential physics. Later qualitatively, we will discuss the role of damping on the stability chart and include it while studying the spatio-temporal dynamics of the BEC.

When the condition $g_{11}g_{22}<g_{12}^{2}$ is satisfied, the condensates are phase segregated. In the opposite limit i.e $g_{11}g_{22}>g_{12}^{2}$, the condensates are in a mixed state. Our study covers the $1D$ case of a cigar shaped BEC extended along the $z$ direction, $V_{i}(\vec r,t)=\dfrac{m}{2}\left[ \Omega_{t_{i}}^{2} (t) (x^{2}+y^{2})+\Omega_{w_{i}}^{2} z^{2}\right]$, where $\Omega_{w_{i}}<<\Omega_{t_{i}}$ is assumed which means that $\Omega_{w_{i}}$ and $\Omega_{t_{i}}$ are the frequencies of the trap for the $i^{th}$ component along the weak and tight confinement directions, respectively. We assume that $\Omega_{t_{i}}$ is subjected to periodic modulation: $\Omega_{t_{i}}(t)=\bar \Omega_{t_{i}}\left[1+\alpha_{i} \cos{\Omega_{i} t} \right]$, $\alpha_{i}<<1$. Here $\Omega_{i}$ is the trap modulation frequency of the $i^{th}$ component. The radial modulation leads to a periodic change of the density of the cloud in time, which is equivalent to a change in the nonlinear interactions and speed of sound. In this paper we will always work either in the deep phase mixed regime or in the deep phase segregated regime. At the boundary separating the phase mixed regime and phase separated regime, the dynamics could be very complex as the trap modulations can continuously switch the system from one regime to the other. In terms of scaled variables, $\tau=\Omega_{12}t$, $\vec R\equiv(X,Y,Z)=(x,y,z)/a_{12}$, $\Omega_{12}=\dfrac{\hbar}{m a_{12}^{2}}$ and $u_{i}=a_{12} \sqrt{4 \pi N_{i} a_{12}} \psi_{i}$, Eqns.(1) and (2) takes the following dimensionless form

\begin{equation}
\imath \dfrac{\partial u_{1}}{\partial t}=\left\lbrace -\dfrac{1}{2}\nabla_{R}^{2}+\dfrac{1}{2}\omega_{t_{1}}^{2}(X^2+Y^2)+\dfrac{1}{2}\omega_{w_{1}}^{2} Z^{2} +\dfrac{g_{1}}{2}|u_{1}|^{2} +|u_{2}|^{2}-\bar \mu_{1} \right\rbrace u_{1} 
\end{equation}

\begin{equation}
\imath \dfrac{\partial u_{2}}{\partial t}=\left\lbrace -\dfrac{1}{2}\nabla_{R}^{2}+\dfrac{1}{2}\omega_{t_{2}}^{2} (X^2+Y^2)+\dfrac{1}{2}\omega_{w_{2}}^{2} Z^{2} +\dfrac{g_{2}}{2}|u_{2}|^{2} +|u_{1}|^{2}-\bar \mu_{2} \right\rbrace u_{2} 
\end{equation},

where $g_{i}=\dfrac{2 g_{ii}}{g_{12}}$, $\omega_{t_{i}}=\dfrac{\Omega_{t_{i}}}{\Omega_{12}}$, $\omega_{w_{i}}=\dfrac{\Omega_{w_{i}}}{\Omega_{12}}$, $\omega_{t_{i}}=\bar \omega_{t_{i}}(1+\alpha_{i} \cos{\omega_{i} \tau})$, $\bar \mu_{i}=\dfrac{\mu_{i}}{\hbar \Omega_{12}}$, $\omega_{i}=\dfrac{\Omega_{i}}{\Omega_{12}}$.

As in \cite{Stalinus04}, we reduce the BEC 3D dynamics to an effective 1D description using a multiple scale analysis \cite{Hasan}. We write the dimensionless wavefunction as:

\begin{equation}
u_{i}=\sqrt{\omega_{i}}     exp\left( {-\dfrac{\omega_{t_{i}}}{2}(X^{2}+Y^{2})}\right)     \phi_{i}(Z,\tau),     i=1,2
\end{equation}

For a flat trap in a weakly confined space $\omega_{w_{i}}=0$ , consequently Eqns. (3) and (4) become

\begin{equation}
2 \imath \dfrac{\partial \phi_{1}}{\partial \tau}=\left\lbrace -\dfrac{\partial^{2}}{\partial Z^{2}}+g_{1} \bar \omega_{t_{1}}|\phi_{1}|^{2}+\omega_{R} |\phi_{2}|^{2}-\mu_{1}'\right\rbrace \phi_{1}+g_{1} \bar \omega_{t_{1}} \alpha_{1} \cos{\omega_{1} \tau}|\phi_{1}|^{2} \phi_{1} 
\end{equation}

\begin{equation}
2 \imath \dfrac{\partial \phi_{2}}{\partial \tau}=\left\lbrace -\dfrac{\partial^{2}}{\partial Z^{2}}+g_{2} \bar \omega_{t_{2}}|\phi_{2}|^{2}+\omega_{R} |\phi_{1}|^{2}-\mu_{2}'\right\rbrace \phi_{2}+g_{2} \bar \omega_{t_{2}} \alpha_{2} \cos{\omega_{2} \tau}|\phi_{2}|^{2} \phi_{2} 
\end{equation},

where $\omega_{R}=\dfrac{2 \omega_{t_{1}}\omega_{t_{2}}}{\omega_{t_1}+\omega_{t_2}}$ and $\mu_{i}'=2 (\bar \mu_{i}-\omega_{t_{i}}), i=1,2$

\begin{figure}[t]
\hspace{-2.0cm}
\includegraphics{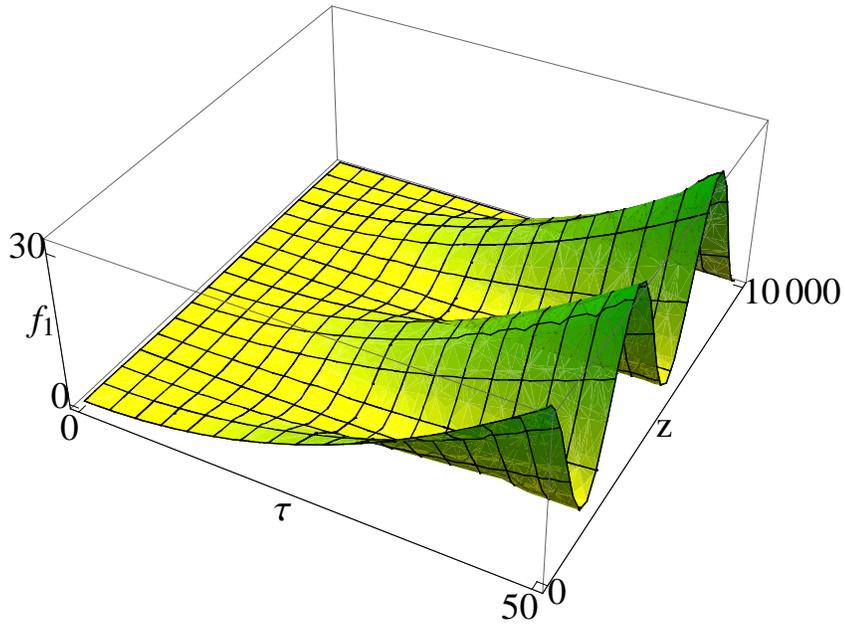} 
\caption{Farday pattern formation in space and time for the first condensate in the phase mixed state. The parameters are $\omega=0.47\times 10^{-3}$, $\alpha=0.1$, $\gamma=0.1$ ,$g_{1}=1.25$ and $g_{2}=1.5$. The time and lenght are in the units of $\Omega_{12}$ and $a_{12}$ respectively.}
\label{3}
\end{figure}
 
\begin{figure}[t]
\hspace{-2.0cm}
\includegraphics{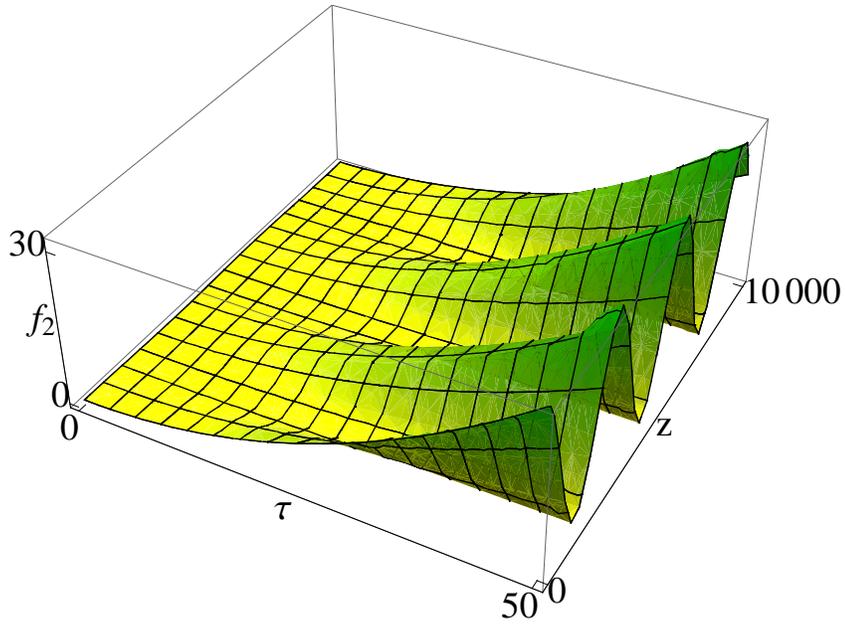} 
\caption{Faraday pattern formation in space and time for the second condensate in the phase mixed state. The parameters are $\omega=0.47\times 10^{-3}$, $\alpha=0.1$, $\gamma=0.1$ ,$g_{1}=1.25$ and $g_{2}=1.5$. The time and lenght are in the units of $\Omega_{12}$ and $a_{12}$ respectively. Note that for the second condensate since the driving frequency is near one of its resonance tongues, the Faraday patterns generated are more compared to that in the first condensate.}
\label{4}
\end{figure}

In order to proceed analytically, we study Eqns. (6) and (7) for the simple case $\alpha_{1}=\alpha_{2}=\alpha$, $\omega_{1}=\omega_{2}=\omega$ and $\omega_{t_{1}}=\omega_{t_{2}}=\omega_{t}$. In this case $\omega_{R}=\bar \omega_{t}(1+\alpha \cos{\omega \tau})$. For a flat potential and in the absence of any perturbation ($\alpha=0$), $\phi_{i}=\phi_{i}^{0}$ is a constant independent of $Z$. Eqns. (6) and (7) yield:

\begin{equation}
|\phi_{1}^{0}|^{2}=\dfrac{\beta_{2}-g_{2} \beta_{1}}{\omega_{t} (1-g_{1}g_{2})}=\delta_{1}
\end{equation}

\begin{equation}
|\phi_{2}^{0}|^{2}=\dfrac{\beta_{1}-g_{1} \beta_{2}}{\omega_{t} (1-g_{1}g_{2})}=\delta_{2}
\end{equation},

where $\beta_{i}=2(\bar \mu_{i}-\bar \omega_{t})$, $i=1,2$. In the presence of the modulations, Eqns. (6) and (7) admits the following homogeneous states:

\begin{equation}
\phi_{1}^{0}=\sqrt{\delta_{1}} exp\left( {-\dfrac{\imath \alpha}{2 \omega} \sin{\omega \tau}}\right) 
\end{equation}

\begin{equation}
\phi_{2}^{0}=\sqrt{\delta_{2}} exp\left( {-\dfrac{\imath \alpha}{2 \omega} \sin{\omega \tau}}\right) 
\end{equation}

This yields, $\mu_{1}'=\mu_{2}'=1$ , $|\phi_{1}^{0}|^{2}=\vert \dfrac{1-g_{2} }{\omega_{t} (1-g_{1}g_{2})}\vert$ and  $|\phi_{2}^{0}|^{2}=\vert \dfrac{1-g_{1} }{\omega_{t} (1-g_{1}g_{2})}\vert$.

Now we wish to know whether the spatially homogeneous external driving fields are able to induce a spontaneous spatial symmetry breaking of this homogeneous state of the coupled BEC's. For that, we perform a linear stability analysis of Eqns. (6) and (7) by adding a small perturbation to Eqn. (10) in the form:

\begin{equation}
\phi_{i}=\phi_{i}^{0}\left( 1+W_{i} \cos{\tilde k_{s_{i}}Z}\right), i=1,2 
\end{equation}

where $W_{i}$ and $\tilde k_{s_{i}}$, $(i=1,2)$ is the complex valued amplitude and the wave-vector of the perturbation of the $i^{th}$ component respectively. Substituting Eqn.(11) into Eqns. (6) and (7) and linearizing with respect to $W_{i}$ leads to the following coupled Mathieu equations for $u_{i}=Re W_{i}$, $(i=1,2)$

\begin{equation}
\dfrac{\partial ^{2}u_{1}}{\partial \tau^{2}}+\left(\Omega_{1}^{2}+\alpha Q_{11} \cos{\omega \tau} \right)u_{1}+Q_{12}\left( 1+\alpha \cos{\omega \tau}\right)u_{2}=0  
\end{equation}

\begin{equation}
\dfrac{\partial ^{2}u_{2}}{\partial \tau^{2}}+\left(\Omega_{2}^{2}+\alpha Q_{22} \cos{\omega \tau} \right)u_{2}+Q_{21}\left( 1+\alpha \cos{\omega \tau}\right)u_{1}=0  
\end{equation},

where $\Omega_{i}(\tilde k_{s_{i}})= \dfrac{\tilde k_{s_{i}}}{2} \sqrt{\tilde k_{s_{i}}^{2}+2q_{ii}}$, is the dispersion relation of the perturbation of the $i^{th}$ condensate in the absence of driving. Also $q_{11}=\vert \dfrac{g_{1}(1-g_{2})}{(1-g_{1}g_{2})} \vert$,  $q_{22}=\vert \dfrac{g_{2}(1-g_{1})}{(1-g_{1}g_{2})} \vert$, $q_{12}=q_{11}/g_{1}$, $q_{21}=q_{22}/g_{2}$, $Q_{11}=\tilde k_{s_{1}}^{2} q_{11}/2$, 
$Q_{22}=\tilde k_{s_{2}}^{2} q_{22}/2$, $Q_{12}=\tilde k_{s_{1}}^{2} q_{12}/2$, $Q_{21}=\tilde k_{s_{2}}^{2} q_{21}/2$.

As known from the general theory of equations of the type (12) and (13), instability may occur for $\omega$ near twice the natural frequencies and their subharmonics, $2 \Omega_{i}/n$, $i=1,2$, $n$ is an integer, and also close to the so-called combination frequencies and their subharmonics, $|\pm \Omega_{1} \pm \Omega_{2}|/n$. The boundary curves for the instability regions must therefore emerge from these frequencies. The wavenumbers $k_{s_{i},n}$ corresponding to the resonance tongues near the natural frequencies and their subharmonics is $k_{s_{i},n}= \sqrt{-q_{ii}+ \sqrt{q_{ii}+4(n \omega/2)^2}}$. In order to find the stability diagrams for the two coupled Mathieu equations, we employ the method indicated in \cite{Hansen}. The method involves infinite determinants which are truncated and then the eigenvalues are found using MATHEMATICA, together with a condition on the characteristic exponent for the solutions on the boundary curves. The result for the phase separated case ($g_{1}=0.5,g_{2}=0.75$) and the phase mixed case ($g_{1}=1.25,g_{2}=1.5$) is shown in Fig.1 and Fig.2 respectively. The shaded regions are the regions of instability. Clearly one notices that the two sets of resonance tongues, namely ${\Omega_{1},\Omega_{2},\dfrac{\Omega_{1}+\Omega_{2}}{2}}$ and ${2\Omega_{1},2\Omega_{2},\Omega_{1}+\Omega_{2}}$ for the phase separated case are more widely spaced with larger stability region in between the tongues than the phase mixed case. The stable regions between the resonance tongues for the phase separated mixture, extends as far as upto $\alpha=0.2$ while those for the phase mixed case extends only upto $\alpha=0.1$. If the detuning of the driving frequency is of the same order for two adjacent modes then these modes can enter in competition. The competition between nearly degenerate modes has been shown to lead to chaotic behaviour in fluids \cite{Cil}. Since the resonance tongues are more closely spaced in the phase mixed state, the mode competition will be stronger for this case.

Now, we study the time and space evolution of the deviation from the initial density $f_{i}=\dfrac{|\phi_{i}|^{2}-|\phi_{i}^{0}|^{2}}{|\phi_{i}^{0}|^{2}}$ by including phenomenologically a damping term proportional to $\gamma \dfrac{d u_{i}}{d \tau}$ in Eqns. (13) and (14). Here $\gamma$ is the damping coefficient. The coupled Mathieu equations (13) and (14) are solved using MATHEMATICA.  Figures (3) and (4) illustrates the time and space evolution of $f_{1}$(density deviation of the first condensate) and $f_{2}$ (density deviation of the second condensate) in the phase mixed state respectively. The parameters are $\omega=0.47\times 10^{-3}$, $\alpha=0.1$, $\gamma=0.1$ , $g_{1}=1.25$ and $g_{2}=1.5$. The periodicity of Faraday patterns generated  is always more for the condensate with lower natural frequency. We found that this observation is also true for the phase separated case. The influence of interactions on the periodicity of pattern formation is clear from the expression $k_{s_{i},n}= \sqrt{-q_{ii}+ \sqrt{q_{ii}+4(n \omega/2)^2}}$. Tuning the interactions, controls the $k_{s_{i}}$ and hence the periodicity.  Also noticed is the fact that onset of pattern formation is not immediate but takes a certain time due to damping. The fact that the phase mixed state is relatively more unstable compared to the phase segregated state was also revealed by the spatio-temporal dynamics (figures not shown). We found that for a given driving $\omega$ and perturbation $\alpha$, the growth rate of the perturbations in the phase mixed state was much higher than that in the phase separated state.   For long enough times in the presence of a continuous modulation, the excitations will grow and eventually the condensates will be destroyed. This contrasts with the everlasting periodic revivals of the spatial modulations in one component 1D case \cite{Abdul}. The spontaneous pattern formation for combinations frequencies is much more complex and simple analytical formulations cannot be done and would require a full numerical simulation of the coupled equations (1) and (2).

\section{Conclusion}

In conclusion, we have demonstrated that the dynamics of Faraday instability in a two component elongated cigar shaped BEC is described by two coupled Mathieu equations. The dynamics depend on the two body intra and inter species interactions. In contrast to the single component case, we found that parametric resonances can occur not only at integral multiples of half the natural frequencies of the two components but also at integral multiples of half the combination frequencies. The stability chart in the $\left\lbrace \omega, \alpha \right\rbrace$ plane reveals that the a phase separated binary mixture is more stable towards pattern formation than the phase mixed mixture. This was confirmed by the spatio-temporal dynamics of the BEC where we found that the initial perturbation grows in a phase mixed state more rapidly than in the phase segregated case.

\end{document}